\documentclass[conference]{IEEEtran}
\IEEEoverridecommandlockouts
\usepackage{cite}
\usepackage{amsmath,amssymb,amsfonts}
\usepackage{algorithmic}
\usepackage{graphicx}
\usepackage{textcomp}
\usepackage{xcolor}
\def\BibTeX{{\rm B\kern-.05em{\sc i\kern-.025em b}\kern-.08em
    T\kern-.1667em\lower.7ex\hbox{E}\kern-.125emX}}
\begin{document}

\title{Simple Framework of ZF Precoding Analysis for Full-Dimensional Massive MIMO Systems\vspace{-5pt}}

\author{Harsh Tataria* and Mansoor Shafi$^\dagger$\\
*Ericsson AB, Lund, Sweden\\
$^{\dagger}$Spark New Zealand, Wellington, New Zealand\\
e-mail: harsh.tataria@ericsson.com and mansoor.shafi@spark.co.nz\vspace{-8pt}}
\maketitle

\begin{abstract}
We provide a simple, yet general and accurate framework for analyzing zero-forcing (ZF) precoding performance for a full-dimensional massive multiple-input multiple-output system. Exploiting an order two Neumann series, our framework approximates the expected (average) ZF signal-to-noise ratio and ergodic sum spectral efficiency, while catering for a finite multipath propagation model, as well as correlated and uncorrelated user equipment (UE) positions in both azimuth and elevation domains. The analysis provides clear insights on the influence of propagation and system parameters on ZF performance. We identify how far UEs must lie in azimuth and elevation for the ZF precoder to approach uncorrelated channels. Our framework is useful for predicting the ZF performance degradation arising from channel correlation. For optimal performance, UEs could be separated by around 30$^{\circ}$ in azimuth and 15$^{\circ}$ in elevation - conditions which are difficult to meet in reality. 
\end{abstract}

\vspace{-12pt}
\section{Introduction}
\label{Introduction}
Massive multiple-input multiple-output (MIMO) is an integral part of fifth-generation (5G) wireless systems. Due to hundreds of electronically steerable antenna elements at the cellular base station (BS), aggressive spatial multiplexing to tens of user equipments (UEs) is possible in the same time-frequency resource \cite{RUSEK1}. For sub-6 GHz massive MIMO operation, digital precoding methods provide a means to suppress multiuser interference. Most current commercial systems employ techniques such as zero-forcing (ZF) precoding to \emph{orthogonalize} multiuser channels, such that scheduled UEs are placed in the null space of one another \cite{RUSEK1,XU1}. Naturally, geographical location of UEs and scatterers play a vital role in conditioning the UE null spaces. If UEs are closely spaced in either azimuth or elevation, ZF suffers from noise enhancement resulting in lower signal-to-noise ratios (SNRs). 

Nonetheless, fundamentally, ``how close is close" in terms of UE positions such that the average ZF SNR to a given UE can be maximized remains to be seen. Naturally, this has a direct consequence on the achievable ergodic sum spectral efficiency over all UEs in the system. Since most studies on massive MIMO (see \cite{NGO1,TATARIA1} and references therein) neglect the presence of the elevation domain, it remains to be seen whether closely separated UEs in the azimuth and elevation domains contribute to the equal ZF performance difference or not, relative to widely separated UEs which result in uncorrelated channels. Important design guidelines on optimal azimuth and elevation separations of scheduled UEs to approach the performance of uncorrelated channels remains to be checked. 

In order to answer these important questions, the focus of this paper is on providing a simple, accurate and general framework for the performance analysis of ZF precoding in full-dimensional \cite{XU1} massive MIMO systems. Despite the vast literature on this topic \cite{RUSEK1,XU1,NGO1,TATARIA1,WANG1}, an insightful framework catering for correlated and uncorrelated finite multipath propagation channels remains missing. To demonstrate its simplicity, our framework utilizes a second-order Neumann series to approximate the expected (average) ZF SNR and ergodic sum spectral efficiency. We provide clear insights on the influence of propagation and system parameters on aforementioned performance metrics. Due to its generality, our framework is capable of handling multiple BS array geometries and estimating the ZF performance degradation as a result of sub-optimal UE positions in the azimuth and elevation. As such, we provide guidelines on the effectiveness of ZF precoding with respect to different UE separations. To our knowledge, a study of this type has been missing from prior literature. 

\vspace{-1pt}
\textbf{Notation.} Upper and lower boldface letters represent matrices and vectors. The $M\times{}M$ identity matrix is denoted as $\mathbf{I}_{M}$, while $M\times{}M$ diagonal matrix $\mathbf{X}$ is referred by $\mathbf{X}=\textrm{diag}(x_{1},x_{2},\dots,x_{M})$. The $(i,j)$-th entry of the matrix $\textbf{X}$ is denoted by $(\mathbf{X})_{i,j}$, while $\mathbf{a}\hspace{1pt}\otimes\hspace{1pt}\mathbf{b}$ denote the Kronecker product of vectors $\mathbf{a}$ and $\mathbf{b}$. Transpose, Hermitian transpose, inverse and trace operators are denoted by $(\cdot)^{T}$, $(\cdot)^{H}$, $(\cdot)^{-1}$ and $\textrm{Tr}[\cdot]$. Moreover, $||\cdot||_{F}$ and  $|\cdot|$ denotes the Frobenius and scalar norms. We use $\mathbf{h}\sim\mathcal{CN}(\mathbf{m},\mathbf{R})$ to denote a complex Gaussian distribution for $\mathbf{h}$ with mean $\mathbf{m}$ and covariance matrix $\mathbf{R}$. Finally, $\mathbb{E}\{\cdot\}$ denotes the statistical expectation.

\vspace{-4pt}
\section{System Model}
\label{SystemModel}
\vspace{-1pt}
We consider the downlink of a single-cell full-dimensional massive MIMO system. The BS is located at the center of a circular cell with a coverage radius of $R_{\textrm{c}}$ meters (m), and is equipped with an array of $M$ transmit antennas. The BS simultaneously serves $L$ single-antenna UEs $\left(M\gg{}L\right)$ in the same time-frequency resource. Uniform power allocation to the $L$ UEs is assumed, and the BS has a constraint on the total transmit effective isotropic radiated power (EIRP) of $P_{\textrm{EIRP}}$ (quoted often in Watts or dB). Knowledge of the channel is assumed (motivated later) with narrowband transmission. The $1\times{}M$ channel from the 
BS array to UE $\ell$ is given by 
\vspace{-4pt}
\begin{equation}
\label{channeltoterminall}
\mathbf{h}_{\ell}=\sum\limits_{p=1}
^{N_{\textrm{P}}}\gamma_{\ell,p}\hspace{2pt}
\mathbf{a}_{\ell,p}\hspace{1pt}.  
\vspace{-5pt}
\end{equation} 
Here $N_{\textrm{P}}$ is the total number of contributing multipath components (MPCs) and 
$\gamma_{\ell,p}\sim\mathcal{CN}
\hspace{2pt}\big(
0\hspace{1pt},1/N_{\textrm{P}}\big)$ is the complex gain on the 
$p$--th MPC. Moreover, and $\mathbf{a}_{\ell,p}$ 
is the $1\times{}M$ far-field array steering vector of the BS array. In the case of a uniform linear array (ULA) positioned on the $z$-axis at the BS, 
\vspace{-3pt}
\begin{align}
\label{ULAsterringvector}
\nonumber
\mathbf{a}_{\ell,p}=
\mathbf{a}\left(\phi_{\ell,p}\right)=&
\left[\hspace{3pt}1\hspace{8pt}e^{-j2\pi{}\frac{d_{z}}
{\lambda_{f}}\cos\hspace{2pt}
\left(\phi_{\ell,p}\right)}\right.\\
&\hspace{20pt}\dots\hspace{4pt}\left.e^{-j2\pi\frac{d_{z}}
{\lambda_{f}}\left(M-1\right)
\cos\hspace{2pt}\left(\phi_{\ell,p}\right)}\right],\\[-20pt]
&\nonumber
\end{align} 
where $d_{z}$ is the inter-element spacing along the $z$-axis of the array, $\lambda_{f}$ is the wavelength associated with carrier frequency, $f$, and $\phi_{\ell,p}$ is the azimuth angle-of-departure (AOD) of the $p$-th MPC. Note that $\gamma_{\ell,p}$, and $\phi_{\ell,p}$ are assumed to be uncorrelated over any of their indices, 
and with each other. Furthermore, $\phi_{\ell,p}\in\left[
\bar{\phi_{\ell}}-\Delta^{\textrm{az}}_{\ell},\bar{\phi_{\ell}}
+\Delta^{\textrm{az}}_{\ell}
\right]$, where $\bar{\phi}_{\ell}$ is the azimuth line-of-sight (LOS) angle to UE 
$\ell$ and $\Delta^{\textrm{az}}_{\ell}$ is the $\ell$-th UEs azimuth angular-spread-of-departure (ASD). In the case when the BS is equipped with a uniform planar array (UPA) on the $x$-$z$ plane, we consider $M_{x}$ antenna elements along the $x$-axis (configured as a ULA), and $M_{z}$ antenna elements along the $z$-axis (also configured as a ULA). The corresponding inter-element separations in $x$ and $z$ directions are denoted by $d_{x}$ and $d_{z}$, respectively. The total number of BS antennas at the UPA is retained as $M=M_{x}M_{z}$. Different from the ULA, the UPA is able to exploit the elevation domain of the channel. To this end, we denote the elevation AOD belonging to the $p$-th MPC from the BS to the $\ell$-th 
UE as $\theta_{\ell,p}$, and note that $\theta_{\ell,p}\in
\left[\bar{\theta}_{\ell}-\Delta^{\textrm{ze}}_{\ell},\bar{\theta}_{\ell}
+\Delta^{\textrm{ze}}_{\ell}\right]$. Here 
$\Delta^{\textrm{ze}}_{\ell}$ is the $\ell$-th UE specific elevation ASD, and $\bar{\theta}_{\ell}$ is the elevation LOS angle to UE $\ell$. In light of the above, the $1\times{}M$ array steering response vector for the $p$-th MPC from the BS to UE $\ell$ is given by  
\begin{equation}
\label{UPAsteeringvector1}
\mathbf{a}_{\ell,p}=
\mathbf{a}\left(\phi_{\ell,p},\theta_{\ell,p}\right)=
\mathbf{a}_{x}\left(\phi_{\ell,p}\hspace{3pt},
\theta_{\ell,p}\right)\otimes
\mathbf{a}_{z}\left(\phi_{\ell,p}\hspace{3pt},
\theta_{\ell,p}\right), 
\end{equation}
where 
\vspace{-1pt}
\begin{align}
\nonumber
\mathbf{a}_{x}\left(\phi_{\ell,p}\hspace{3pt},
\theta_{\ell,p}\right)=&\left[\hspace{2pt}1\hspace{8pt}
e^{-j2\pi\frac{d_{x}}{\lambda_{f}}\sin\hspace{1pt}
\left(\theta_{\ell,p}\right)
\cos\hspace{1pt}\left(\phi_{\ell,p}\right)}\hspace{3pt}\right.\\
\label{UPAsteeringvector2}
&\dots\hspace{3pt}\left.e^{-j2\pi\frac{d_{x}}{\lambda_{f}}
\left(M_{x}-1\right)\sin\hspace{1pt}
\left(\theta_{\ell,p}\right)
\cos\hspace{1pt}\left(\phi_{\ell,p}\right)}\right],\\[-25pt]
&\nonumber
\end{align}
and 
\begin{align}
\nonumber
\mathbf{a}_{z}\left(\phi_{\ell,p}\hspace{3pt},
\theta_{\ell,p}\right)=&\left[\hspace{1pt}1\hspace{8pt}
e^{-j2\pi\frac{d_{z}}{\lambda_{f}}\sin\hspace{1pt}
\left(\theta_{\ell,p}\right)
\sin\hspace{1pt}\left(\phi_{\ell,p}\right)}\right.\\
\label{UPAsteeringvector3}
&\left.\hspace{3pt}\dots
\hspace{3pt}e^{-j2\pi\frac{d_{z}}{\lambda_{f}}
\left(M_{z}-1\right)\sin\hspace{1pt}
\left(\theta_{\ell,p}\right)
\sin\hspace{1pt}\left(\phi_{\ell,p}\right)}\right]. \\[-18pt]
&\nonumber
\end{align}
For maximum clarity, we defer the discussion of the aforementioned parameters till Sec.~\ref{NumericalResults}. 

\textbf{Remark 1.} Two reasons can be pointed out to justify the rather impractical assumption of perfect channel knowledge. Firstly, unlike previous studies, the central focus of the work is to devise a simple, yet accurate, analysis framework to gain insights into the behavior of expected ZF SNR and ergodic sum spectral efficiency when considering full-dimensional massive MIMO systems. In contrast to prior studies, spatial channel parameters in both azimuth and zenith domains are utilized to capture their variations across multiple terminals. Under such a heterogeneous scenario, it is extremely difficult to make analytical progress without perfect channel knowledge.  Secondly, it is worth noting that the results obtained from the subsequent analysis and evaluations can be treated as a useful upper bound on the performance which may be seen in practice with estimated channels. 

The received signal at UE $\ell$ is given by 
\begin{equation}
    \label{RXSignalUEl}
    y_{\ell}=\sqrt{\frac{P_{\textrm{EIRP}}\hspace{1pt}\beta_\ell}{\eta}}\hspace{1pt}\mathbf{h}_{\ell}\hspace{1pt}\mathbf{g}_\ell{}\hspace{1pt}s_\ell+\sum\limits_{\substack{i=1\\i\neq{}\ell}}^{L}\sqrt{\frac{P_{\textrm{EIRP}}\hspace{1pt}\beta_\ell}{\eta}}\hspace{1pt}\mathbf{h}_\ell\hspace{1pt}\mathbf{g}_{i}\hspace{1pt}s_{i}+n_\ell, 
    \vspace{-2pt}
\end{equation}
where $\beta_\ell$ denotes the link gain of UE $\ell$ composing of the large-scale propagation effects of pathloss and shadow fading. Parameters involved in $\beta_\ell$ are discussed in Sec.~\ref{NumericalResults}. Additionally, $\mathbf{g}_\ell$ is the $M\times{}1$ un-normalized precoding vector from the BS to UE $\ell$, obtained from the $\ell$-th column of $\mathbf{G}$, the composite un-normalized $M\times{}L$ precoding matrix of all $L$ UEs. The data symbol for UE $\ell$ is given by $s_\ell$, s.t. $\mathbb{E}\{|s_\ell|^2\}=1$, $\forall\ell=\{1,2,\dots,L\}$. Finally, the additive white Gaussian noise term at UE $\ell$ is denoted by $n_\ell$, such that $n_\ell\sim\mathcal{CN}(0,\sigma^2)$. For simplicity, we fix $\sigma^2$ $\forall{}\ell=\{1,2,\dots,L\}$. The precoder normalization parameter is given by $\eta=||\mathbf{G}||^2_{F}/L$, s.t. $\mathbb{E}\{||\mathbf{g}_\ell||^2\}=1,\forall\ell=\{1,2,\dots,L\}$. 

In the case of ZF precoding, $\mathbf{g}_\ell$ forms the $\ell$-th column of $\mathbf{G}=\mathbf{H}^H(\mathbf{HH}^H)^{-1}$, where $\mathbf{H}=[\mathbf{h}_1^T, \mathbf{h}_2^T,\dots,\mathbf{h}_L^T]^T$ is the $L\times{}M$ matrix of all $L$ UE channels. Noting that $\mathbf{HG}=\mathbf{HH}^{H}(\mathbf{HH}^{H})^{-1}=\mathbf{I}_{L}$, the ZF SNR for UE $\ell$ is given by \cite{TATARIA1} 
\begin{equation}
\label{zfsnrterminall}
\textrm{SNR}_{\ell}=\frac{P_{\textrm{EIRP}}\hspace{1pt}
\beta_{\ell}}{\sigma^{2}\hspace{1pt}\eta}=\frac{P_{\textrm{EIRP}}\hspace{1pt}
\beta_{\ell}}
{\sigma^{2}\left\{\frac{1}{L}\left\{\textrm{Tr}\left[\left(\mathbf{HH}^{H}
\right)^{-1}\hspace{1pt}
\right]\hspace{1pt}\right\}\hspace{1pt}\right\}}, 
\end{equation}
with $\eta=\|\mathbf{G}\|^{2}_{F}/L=\textrm{Tr}\hspace{2pt}
[\hspace{1pt}(\hspace{1pt}\mathbf{HH}^{H})^{-1}\hspace{1pt}]/L$. The expression in \eqref{zfsnrterminall} can be readily used to derive the ergodic sum  spectral efficiency (in bits/s/Hz) over all $L$ UEs. This is given by 
\vspace{-3pt}
\begin{equation}
    \label{ergodicsumspectralefficiency}
    \textrm{R}=\mathbb{E}\left\{\hspace{1pt}
    \sum\limits_{\ell=1}^{L}\hspace{2pt}\log_2\left(1+\textrm{SNR}_{\ell}\right)
    \right\}, 
    \vspace{-2pt}
\end{equation}
with the expectation performed over small-scale fading in $\mathbf{H}$. 

\textbf{Remark 2.} Analyzing the ergodic sum spectral efficiency performance with ZF precoding in full-dimensional massive MIMO systems is a difficult task. This is since finding \emph{exact} moments of the underlaying ZF SNR in \eqref{zfsnrterminall} is challenging due to the matrix trace in the denominator of \eqref{zfsnrterminall} being a random function of the inverse of the composite channel correlation matrix.  Moreover, the structure of $\mathbf{H}$ is fully heterogeneous, involving both azimuth and elevation components. Below we devise a simple, yet accurate, analysis framework of the ZF ergodic sum spectral efficiency via the expected ZF SNR.

\vspace{-1pt}
\section{Analysis Framework and Insights}
\label{AnalysisFrameworkandImplications}
\vspace{-1pt}
As exact analysis of ZF SNR moments is difficult, we \emph{approximate} the inverse in \eqref{zfsnrterminall} with a finite order Neumann series expansion \cite{TATARIA1,WANG1}. We first express $\mathbf{HH}^{H}$ as 
\vspace{-3pt}
\begin{equation}
\label{zfanalysis1}
\mathbf{HH}^{H}=\mathbb{E}\left\{\mathbf{HH}^{H}
\right\}+\mathbf{\Xi}\hspace{1pt}=\mathbf{\Delta}+\mathbf{\Xi},
\vspace{-5pt}
\end{equation}
where $\mathbf{\Xi}=\mathbf{HH}^{H}-\mathbf{\Delta}$ 
and $\mathbb{E}\left\{
\mathbf{\Xi}\right\}$ may not be zero. This is in sharp contrast to the analysis carried out in \cite{TATARIA1} for simple Rayleigh fading channels, where $\mathbb{E}\{\mathbf{\Delta}\}=\mathbb{\mathbf{\Xi}}=0$. With an order $N$ Neumann series, the inverse of \eqref{zfanalysis1} can be written by
\vspace{-15pt}
\begin{align}
\nonumber
\left(\mathbf{HH}^{H}\right)^{-1}=&
\left(\mathbf{\Delta}+\mathbf{\Xi}\hspace{1pt}\right)^{-1}=
\left[\mathbf{\Delta}\left(\mathbf{I}_{L}+
\mathbf{\Delta}^{-1}\hspace{2pt}\mathbf{\Xi}
\hspace{1pt}\right)\hspace{1pt}\right]^{-1}\\[-5pt]
\approx&\sum\limits_{n=0}^{N}\hspace{1pt}
\left(\hspace{1pt}-\mathbf{\Delta}^{-1}\hspace{2pt}
\mathbf{\Xi}\hspace{1pt}
\right)^{n}\mathbf{\Delta}^{-1}. 
\label{zfanalysis2}\\[-19pt]
&\nonumber
\end{align}
Substituting the definition of $\mathbf{\Xi}$, we can write 
\eqref{zfanalysis2} as 
\vspace{-3pt}
\begin{equation}
\label{zfanalysis3}
\left(\mathbf{HH}^{H}\right)^{-1}\approx
\sum\limits_{n=0}^{N}\hspace{3pt}\left[\hspace{1pt}
-\hspace{1pt}\mathbf{\Delta}^{-1}
\left(\mathbf{HH}^{H}-\mathbf{\Delta}\right)
\hspace{1pt}\right]^{n}\mathbf{\Delta}^{-1}.
\vspace{-3pt}
\end{equation}
Analyzing the Neumann series in \eqref{zfanalysis3}, when $N=2$ gives the expressions of \eqref{zfanalysis4} (see top of the following page for space reasons)\footnote{Knowing the series order apriori is a rather difficult task, which in general does not have a prescribed methodology. For most common channel descriptions (see e.g.\cite{WANG1}), expansion with $N=2$ has shown to converge to the true inverse of $\mathbf{HH}^{H}$. Using this as our motivation, we also employ $N=2$ in this paper to develop the analysis framework.}. 
\begin{figure*}[!t]
\begin{align}
\nonumber
\left(\mathbf{HH}^{H}\right)^{-1}\hspace{-3pt}
\approx&\hspace{4pt}
\mathbf{\Delta}^{-1}-\left[\hspace{1pt}
\mathbf{\Delta}^{-1}
\left(\mathbf{HH}^{H}-\mathbf{\Delta}\right)
\mathbf{\Delta}^{-1}\right]+\left[\hspace{1pt}\mathbf{\Delta}^{-1}
\left(\mathbf{HH}^{H}-\mathbf{\Delta}\right)
\mathbf{\Delta}^{-1}
\left(\mathbf{HH}^{H}-\mathbf{\Delta}\right)
\mathbf{\Delta}^{-1}\right]\\[6pt]
\label{zfanalysis4}
=&\hspace{3pt}3\hspace{1pt}
\mathbf{\Delta}^{-1}-3\hspace{1pt}
\mathbf{\Delta}^{-1}\hspace{1pt}
\mathbf{HH}^{H}\hspace{1pt}
\mathbf{\Delta}^{-1}+\mathbf{\Delta}^{-1}\hspace{1pt}\mathbf{HH}^{H}
\mathbf{\Delta}^{-1}\hspace{1pt}\mathbf{HH}^{H}\mathbf{\Delta}^{-1}. \\[-25pt]
&\nonumber
\end{align}
\vspace{-5pt}
\hrulefill
\end{figure*}
Defining $\mathbf{Z}=\mathbf{\Delta}^{-1}\mathbf{HH}^H\mathbf{\Delta}^{-1}$ and substituting the resulting expression from \eqref{zfanalysis4} into \eqref{zfsnrterminall} yields 
\begin{equation}
\label{zfanalysis5}
\textrm{SNR}_{\ell}\approx\frac{P_{\textrm{EIRP}}\hspace{2pt}\beta_{\ell}}{\sigma^{2}\bigg\{
\frac{1}{L}\Big\{\hspace{1pt}
\textrm{Tr}\hspace{1pt}
\big[\hspace{1pt}3\mathbf{\Delta}^{-1}\hspace{-2pt}-3\hspace{1pt}
\mathbf{Z}\hspace{-1pt}+\hspace{-1pt}\mathbf{Z}\hspace{1pt}\mathbf{HH}^{H}\hspace{-2pt}\mathbf{\Delta}^{-1}
\hspace{1pt}\big]\hspace{2pt}\Big\}\hspace{1pt}\bigg\}}\hspace{1pt}.
\vspace{-2pt}
\end{equation}
After further algebraic manipulations, the matrix trace in the denominator of \eqref{zfanalysis5} 
can be broken down into the trace of the involved terms \cite{HJ1}. This is given by
\vspace{-4pt}
\begin{align}
\nonumber
\textrm{Tr}\left[\hspace{1pt}
\left(\hspace{1pt}\mathbf{HH}^{H}\hspace{1pt}\right)^{-1}\right]\approx&\hspace{3pt}
3\hspace{2pt}\textrm{Tr}\left[\hspace{1pt}
\mathbf{\Delta}^{-1}\hspace{1pt}\right]-3\hspace{2pt}
\textrm{Tr}\left[\hspace{1pt}\mathbf{HH}^{H}\mathbf{\Delta}^{-2}\hspace{1pt}\right]\\
\label{zfanalysis6}
&\hspace{-1pt}+\textrm{Tr}\left[\hspace{1pt}\mathbf{HH}^{H}\mathbf{\Delta}^{-1}\hspace{1pt}\mathbf{HH}^{H}\mathbf{\Delta}^{-2}\hspace{1pt}\right]. \\[-21pt]
&\nonumber
\end{align}

\textbf{Remark 3.} In what follows, we evaluate the expected value of \eqref{zfanalysis5}, using the result in \eqref{zfanalysis6}. To overcome the cumbersome nature of the expectation over small-scale fading, we employ the univariate special case of the commonly used Laplace approximation \cite{TATARIA1}. This allows us to write \eqref{zfanalysis7} (on top of the following page due to space constraints). \begin{figure*}[!t]
\begin{equation}
\label{zfanalysis7}
\mathbb{E}\left\{\textrm{SNR}_{\ell}\right\}
\approx\frac{P_{\textrm{EIRP}}\hspace{2pt}
\beta_{\ell}}{\sigma^{2}\hspace{2pt}\bigg\{\hspace{2pt}
\frac{1}{L}\hspace{2pt}\bigg\{
\mathbb{E}\left\{3\hspace{2pt}\textrm{Tr}\left[\hspace{1pt}
\mathbf{\Delta}^{-1}\right]-3\hspace{2pt}
\textrm{Tr}\left[\mathbf{HH}^{H}\mathbf{\Delta}^{-2}
\right]+\textrm{Tr}\left[\mathbf{HH}^{H}\mathbf{\Delta}^{-1}
\hspace{1pt}\mathbf{HH}^{H}\mathbf{\Delta}^{-2}\right]
\hspace{2pt}\bigg\}\hspace{3pt}\bigg\}\hspace{2pt}\right\}}. 
\end{equation}
\hrulefill
\vspace{-14pt}
\end{figure*}
The approximation introduced in \eqref{zfanalysis7} is of the form $\mathbb{E}\{\delta/X\}=\delta/
\mathbb{E}\{X\}$, where $\delta$ is a scalar value. The accuracy of such approximations rely on $X$ having a small variance compared to its mean value - an effect which can be observed by applying a Taylor series expansion to $\delta/X$ around $\delta/\mathbb{E}\{X\}$. The terms in \eqref{zfanalysis7} are well suited to this approximation, especially when $M$ and $L$ start to grow (common for a massive MIMO system), since the implicit averaging in the denominator of \eqref{zfanalysis7} gives rise to the variance reduction required for convergence (see e.g., \cite{TATARIA1}). 

We further analyze \eqref{zfanalysis7} to derive an analytical solution with an order two Neumann series. Taking the expectation through the trace in the denominator of \eqref{zfanalysis7} allows us to write its three individual terms as 
\vspace{-5pt}
\begin{equation}
    \label{term1zfanalysis8}
    T_{1}=3\hspace{2pt}\textrm{Tr}\left[\hspace{1pt}\mathbb{E}\hspace{1pt}\{\mathbf{\Delta}^{-1}\}\hspace{1pt}\right], 
    \vspace{-3pt}
\end{equation}
\vspace{-3pt}
\begin{equation}
    \label{term2zfanalysis8}
    T_{2}=3\hspace{2pt}\textrm{Tr}\left[\hspace{1pt}\mathbb{E}\left\{\mathbf{HH}^{H}\mathbf{\Delta}^{-2}\right\}\hspace{1pt}\right], 
    \vspace{-2pt}
\end{equation}
and 
\vspace{-2pt}
\begin{equation}
    \label{term3zfanalysis8}
    T_{3}=\textrm{Tr}\hspace{1pt}\left[\hspace{1pt}    \mathbb{E}\left\{\mathbf{HH}^{H}\mathbf{\Delta}^{-1}\hspace{2pt}\mathbf{HH}^{H}\mathbf{\Delta}^{-2}\right\}\hspace{1pt}\right],  
    \vspace{-2pt}
\end{equation}
respectively. As $\mathbb{E}\{\mathbf{HH}^{H}\}=\mathbf{\Delta}$, the first two terms result in a cancellation, allowing us to express  \eqref{zfanalysis7} as 
\vspace{-1pt}
\begin{align}
\nonumber
\mathbb{E}\left\{\textrm{SNR}_{\ell}\right\}\approx&\\[-3pt]
&\hspace{-40pt}\frac{P_{\textrm{EIRP}}\hspace{2pt}\beta_{\ell}}
{\sigma^{2}\hspace{1pt}\bigg\{\frac{1}{L}\hspace{1pt}
\bigg\{\hspace{1pt}\textrm{Tr}\left[\hspace{1pt}
\mathbb{E}\left\{
\mathbf{HH}^{H}\mathbf{\Delta}^{-1}
\hspace{1pt}\mathbf{HH}^{H}\right\}
\mathbf{\Delta}^{-2}\hspace{2pt}
\right]\bigg\}\hspace{3pt}\bigg\}}\hspace{1pt}. 
\label{zfanalysis9}
\end{align}
We now evaluate the remaining expectation in the denominator of \eqref{zfanalysis9}. Separating the expectation into its respective terms, for any $\ell\neq{}j\in\left\{1,2,\dots,
L\right\}$ we can write  
\vspace{-1pt}
\begin{align}
\nonumber
\mathbb{E}\left\{\hspace{1pt}\left(\hspace{1pt}\mathbf{HH}^{H}
\mathbf{\Delta}^{-1}\hspace{1pt}\mathbf{HH}^{H}\hspace{1pt}\right)_{\ell,j}\right\}\hspace{-1pt}
\overset{(a)}=&\hspace{2pt}
\mathbb{E}\left\{\hspace{2pt}\mathbf{h}_{\ell}\left(\mathbf{\Psi}\right)\mathbf{\Delta}^{-1}\left(\mathbf{\Psi}
\right)^{H}\mathbf{h}_{j}^{H}\right\}\\[3pt]
\nonumber
&\hspace{-45pt}=\mathbb{E}\left\{
\sum\limits_{r=1}^{L}\sum\limits_{s=1}^{L}
\hspace{2pt}
\mathbf{h}_{\ell}\hspace{2pt}
\mathbf{h}_{r}^{H}\left(\mathbf{\Delta}^{-1}\right)_{\hspace{-1pt}
r,s}\hspace{-2pt}\mathbf{h}_{s}
\hspace{2pt}\mathbf{h}_{j}^{H}
\right\}\\[3pt]
\nonumber
&\hspace{-45pt}\overset{(b)}=\hspace{2pt}\sum\limits_{r=1}^{L}
\sum\limits_{s=1}^{L}\mathbb{E}\left\{
\mathbf{h}_{\ell}\hspace{2pt}
\mathbf{h}_{r}^{H}\hspace{1pt}
\mathbf{h}_{s}\hspace{2pt}
\mathbf{h}_{j}^{H}\right\}
\left(\mathbf{\Delta}^{-1}\right)_{r,s}\\[6pt]
\label{zfanalysis10}
&\hspace{-45pt}=M^{2}\left(\mathbf{\Delta}^{-1}\right)_{\ell,j}, 
\end{align}
where $(a)$ contains $\mathbf{\Psi}=(\mathbf{h}_{1}^{H}\hspace{2pt}
\mathbf{h}_{2}^{H}\hspace{2pt}\dots\hspace{2pt}\mathbf{h}_{L}^{H})$, while $(b)$ contains straightforward manipulations to arrive at the result in \eqref{zfanalysis10}. Likewise, when $\ell=j\in\{1,2,\dots,L\}$, we can express 
\begin{align}
\nonumber
\mathbb{E}\left\{
\left(\mathbf{HH}^{H}
\mathbf{\Delta}^{-1}\hspace{1pt}\mathbf{HH}^{H}\right)_{\ell,\ell}
\right\}\hspace{-1pt}=\mathbb{E}\left\{\mathbf{h}_{\ell}
\hspace{1pt}
\mathbf{h}_{\ell}^{H}\hspace{1pt}
\mathbf{h}_{\ell}\hspace{1pt}
\mathbf{h}_{\ell}^{H}\right\}\left(
\mathbf{\Delta}^{-1}\right)_{\ell,\ell}\\ 
\nonumber
&\hspace{-150pt}
+\sum\limits_{\substack{r=1\\r\neq{}\ell}}^{L}
\mathbb{E}\left\{\mathbf{h}_{\ell}\hspace{2pt}\mathbf{h}_{r}^{H}
\mathbf{h}_{r}\hspace{1pt}\mathbf{h}_{\ell}^{H}\right\}
\left(\mathbf{\Delta}^{-1}\right)_{r,r}\\ \nonumber
&\hspace{-150pt}=\left(\mathbf{\Delta}^{-1}
\right)_{\ell,\ell}\hspace{1pt}\mathbb{E}\left\{
\mathbf{h}_{\ell}\hspace{2pt}
\mathbf{h}_{\ell}^{H}\hspace{2pt}
\mathbf{h}_{\ell}\hspace{2pt}
\mathbf{h}_{\ell}^{H}
\right\}\\ 
\label{zfanalysis11}
&\hspace{-150pt}+\sum\limits_{\substack{r=1\\r\neq{}\ell}}^{L}
\left(\mathbf{\Delta}^{-1}\right)_{r,r}\textrm{Tr}
\left[\hspace{1pt}\mathbf{R}_{r}\mathbf{R}_{\ell}\hspace{1pt}\right],\\[-20pt]
&\nonumber
\end{align}
where $\mathbf{R}_{\ell}=\mathbb{E}\{\mathbf{h}_{\ell}\hspace{1pt}\mathbf{h}_{\ell}^{H}\}$. Recognizing that the expectation in \eqref{zfanalysis11} can be evaluated by stating $\mathbb{E}\left\{\mathbf{h}_{\ell}\hspace{1pt}\mathbf{h}_{\ell}^{H}\hspace{1pt}
\mathbf{h}_{\ell}\hspace{1pt}
\mathbf{h}_{\ell}^{H}\right\}=\mathbb{E}\left\{|\hspace{1pt}\mathbf{h}_{\ell}
\hspace{1pt}\mathbf{h}_{\ell}^{H}|^{2}\right\}$, following lengthy but straightforward calculations result in $\mathbb{E}\left\{
|\mathbf{h}_{\ell}\hspace{1pt}\mathbf{h}_{\ell}^{H}|^{2}\right\}=M^{2}+\textrm{Tr}\hspace{2pt}[\mathbf{R}_{\ell}^{2}]$. We only resort to only a brief presentation due to space reasons. As such, 
\begin{align}
\nonumber
\mathbb{E}\left\{\hspace{2pt}
\left(\hspace{1pt}\mathbf{HH}^{H}
\mathbf{\Delta}^{-1}\mathbf{HH}^{H}\right)_{\ell,\ell}
\right\}=&\left(\mathbf{\Delta}^{-1}\right)_{\ell,\ell}
\Big\{M^{2}+\textrm{Tr}\left[\mathbf{R}_{\ell}^{2}
\right]\hspace{-1pt}\Big\}\hspace{2pt}\\
&\hspace{-45pt}+\sum\limits_{\substack{r=1\\r\neq{}\ell}}
^{L}\left(\mathbf{\Delta}^{-1}\right)_{r,r}
\Big\{\textrm{Tr}\left[\mathbf{R}_{r}\hspace{1pt}\mathbf{R}_{\ell}\hspace{1pt}\right]
\Big\}.\label{zfanalysis12}\\[-20pt]
&\nonumber
\end{align}
Combining the results in \eqref{zfanalysis12} and \eqref{zfanalysis10}, the overall expectation  
\begin{align}
\nonumber
\mathbb{E}\left\{\mathbf{HH}^{H}\mathbf{\Delta}^{-1}
\mathbf{HH}^{H}\right\}&=M^{2}\mathbf{\Delta}^{-1}\hspace{-1pt}+\hspace{-1pt}\textrm{diag}\left(d_{1},d_{2},\dots,d_{L}\right)\\ \label{zfanalysis13}
&=M^{2}\mathbf{\Delta}^{-1}
+\mathbf{D}, \\[-20pt]
&\nonumber
\end{align}
where $d_{\ell}=\sum\nolimits_{r=1}^{L}
(\mathbf{\Delta}^{-1})_{\hspace{2pt}r,r}\hspace{2pt}
\textrm{Tr}\hspace{2pt}[\mathbf{R}_{r}
\mathbf{R}_{\ell}]$. Substituting \eqref{zfanalysis13} into \eqref{zfanalysis9}, we obtain $\mathbb{E}\{\textrm{SNR}_\ell\}=\mathbb{E}\{\textrm{SNR}_{\ell,\textrm{approx}}\}$ as
\begin{align}
\nonumber
\hspace{-25pt}\mathbb{E}\left\{\textrm{SNR}_{\ell,\textrm{approx}}\right\}\approx&\\[-3pt]
\label{zfanalysis14}
&\hspace{-40pt}\frac{P_{\textrm{EIRP}}\hspace{2pt}\beta_{\ell}}{\sigma^{2}\hspace{1pt}\bigg\{\frac{1}{L}\hspace{1pt}\bigg\{\hspace{1pt}\textrm{Tr}\left[\hspace{2pt}\left(M^{2}\mathbf{\Delta}^{-1}+\mathbf{D}\right)\mathbf{\Delta}^{-2}\hspace{1pt}\right]\bigg\}\bigg\}}.
\end{align}
This can be used to directly approximate the ergodic sum spectral efficiency by 
\vspace{-4pt}
\begin{equation}
    \label{zfanalysis15}
    \textrm{R}_{\textrm{approx}}\approx{}L\log_2
    \left(1+\mathbb{E}\left\{\textrm{SNR}_{\ell,\textrm{approx}}\right\}\right). 
    \vspace{-3pt}
\end{equation}
While the tightness of \eqref{zfanalysis14} and \eqref{zfanalysis15} is assessed and discussed in Sec. \ref{NumericalResults} relative to their simulated counterparts, below we give some remarks to draw insights on the derived results. 

\textbf{Remark 4.} The results in \eqref{zfanalysis14} and \eqref{zfanalysis15} present a simple solution to a difficult problem, lending itself to some useful insights. The denominator of \eqref{zfanalysis14} (expected noise power of UE $\ell$) is primarily governed by $\textrm{Tr}[\mathbf{\Delta}]=\textrm{Tr}[\hspace{1pt}\mathbb{E}\{\mathbf{HH}^{H}\}\hspace{1pt}]$ and $\textrm{Tr}[\mathbf{R}_{\ell}\mathbf{R}_{r}]$. Maximization of the both trace terms would lead to maximum expected noise power. This happens when $\mathbf{R}_{\ell}$ starts to become similar to $\mathbf{R}_{r}$, i.e., the \emph{maximum eigenvectors} of $\mathbf{R}_{\ell}$ and $\mathbf{R}_{r}$ are aligned/in-phase. In general, such an alignment is caused by similarity of channels to UEs $\ell$ and $r$ where $r\neq{}\ell$. This is in turn controlled by the propagation channel's spatial parameters such as the LOS angles and angular spreads in both azimuth and elevation domains. In general, non-overlapping angular support sets would result in independent channels, meaning that the UE and scatterer locations do not correlate the individual channel responses. Typically, when the UEs are closely spaced, their impulse responses are highly correlated, where the $\textrm{Tr}[\hspace{1pt}\mathbb{E}\{\mathbf{HH}^{H}\}\hspace{1pt}]$ and $\textrm{Tr}[\mathbf{R}_{\ell}\hspace{1pt}\mathbf{R}_{r}]$ approach their maximum value. Later we demonstrate this effect qualitatively by experimenting with the relative UE angular separations in both azimuth and elevation domains. Furthermore, another prominent factor influencing the maximization of expected noise power is the ratio of $M^2/L$, implying that for a given number of BS antennas, increasing the number of UEs will \emph{linearly} increase the expected noise power and thus decrease the overall ZF SNR. Such insights are difficult to obtain from more complex solutions derived in the literature (see e.g., \cite{ADHIKARY1}), for simpler propagation channels which require a linked set of equations, even in the large system regime. In sharp contrast, our analysis poses no such constraints and is general to the operational system dimension, channel model, channel correlation structure and antenna array geometries.

\vspace{-3pt}
\section{Numerical Results}
\label{NumericalResults}
We consider an urban non-LOS macrocellular environment from the propagation channel model in \cite{3GPPTR38901}. Operating at 3.7 GHz ($\lambda_f$=8.1 cm), we have $R_{c}$ = 500 m with $\sigma^2=1;\hspace{3pt}\forall{}\ell=\{1,2,\dots{},L\}$. As such we refer to the term ``operating SNR" to denote $P_{\textrm{EIRP}}/\sigma^2$. We employ inter-element spacing of 0.5$\lambda_f$ in the azimuth and 0.7$\lambda_f$ in elevation. The link gain for UE $\ell$, $\beta_\ell$, follows the model quoted in Table 7.4.1-1 of \cite{3GPPTR38901} with pathloss exponent of 3.67 and zero-mean log-normally distributed shadowing having a standard deviation of 4 dB. The total number of MPCs, $N_P$, and other parameters such as the ASDs in azimuth and elevation are employed from Table 7.5-6 Part I in \cite{3GPPTR38901}. Unless otherwise stated, we assume that the BS is equipped with UPA having $M=128$ antennas (16 rows, 8 columns) serving $L=4$ UEs. This is since the UPA better captures the full-dimensional channel, relative to the ULA. Though this is our default system configuration, we confirm that the overall trends do not change with varying values of $M$ or $L$. Inter-UE angular separations in the azimuth and elevation are defined in relation to a given UE's azimuth and elevation LOS angles. Relative to these, a constant offset in the LOS angles determines the angular separation of other UEs. This allows us to study the impact of closely spaced UEs relative to UEs which may be far apart. We first evaluate the accuracy of the proposed Neumann series approximation, followed by expected SNR and ergodic sum spectral efficiency estimates for a range of UE placements. 

\begin{figure}[!t]
    \centering
    \includegraphics[width=8.1cm]{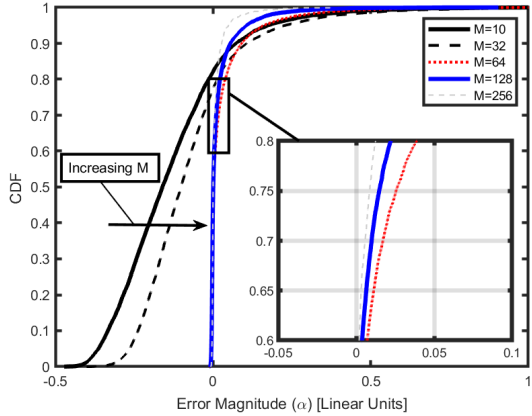}
    \vspace{-8pt}
   \caption{CDFs of Neumann series approximation error magnitude as a function of BS antenna numbers when $N=2$, $L=4$ and operating SNR = 0 dB.}
\label{fig:NSAccuracy}
\vspace{-15pt}
\end{figure}

Figure~\ref{fig:NSAccuracy} depicts the Neumann series approximation accuracy assessed by comparing the true $(\mathbf{HH}^H)^{-1}$ relative to that derived in \eqref{zfanalysis4}. In order to quantify the total approximation accuracy, with $N=2$, we define the following metric: $\textrm{error magnitude},\alpha=\textrm{Tr}[\hspace{1pt}\mathbf{I}_{L}-[(\mathbf{HH}^H)^{-1}/\hspace{1pt}\textrm{approx.NS}\hspace{1pt}]\hspace{1pt}]$, where ``approx.NS" represents \eqref{zfanalysis4}. The figure shows the cumulative distribution functions (CDFs) of the error magnitude with variation in the number of BS antennas, $M$ with $L=4$ at operating SNR = 0 dB. Two interesting trends can be observed: Firstly, increasing $M$ significantly reduces the error magnitude (comparing when $M$=10 and 32 to $M$=64). This difference is due to the improvement brought by scaling up the degrees-of-freedom induced by higher values of $M$ relative to $L$. At $M=64$ the error magnitude CDF seems to have largely converged to zero, while the top end of the CDF still exhibits very small error magnitudes. Secondly, errors at the top end of the CDF start to disappear as $M$ is increased from 64 to 128 and 256, as seen by the zoomed part of the figure, where the inverse becomes well conditioned. Though not shown here due to space reasons, we confirm that if the \emph{order} of the Neumann series is reduced to $N=1$, the approximation suffers from 40\% loss in accuracy around CDF = 0.5. 

\begin{figure}[!t]
    \centering
    \includegraphics[width=8.1cm]{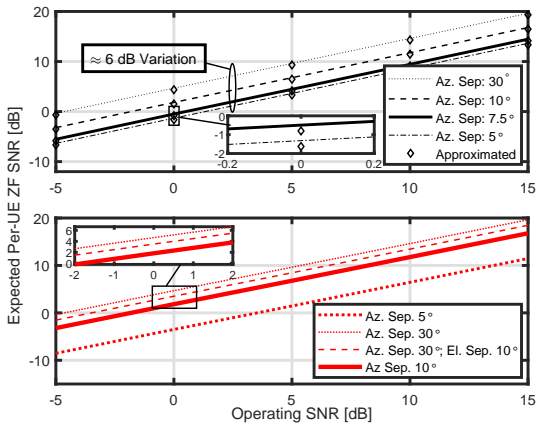}
    \vspace{-10pt}
   \caption{Expected per-UE ZF SNR vs. operating SNRs with $M$=128, $L$=4, and varying azimuth and elevation UE separations.}
    \label{fig:ExpectedZFSNRvsSNR}
    \vspace{-18pt}
\end{figure}

Figure~\ref{fig:ExpectedZFSNRvsSNR} shows the expected per-UE ZF SNR as a function of operating SNR with $M$=128, $L$=4 for various inter-UE separations in both azimuth and elevation domains. Some important conclusions can be drawn from the figure. Firstly, in the top sub-figure, the impact of UE separation in the azimuth domain is examined. For consistency in comparison, elevation UE separation of 15$^\circ$ and azimuth UE separation of 30$^\circ$ denotes uncorrelated channels. As the UE separation is reduced, one can notice a significant reduction (up to 6 dB) in the expected ZF SNR. This is seen by comparing azimuth UE separations of 30$^\circ$ relative to 7.5$^\circ$ or 5$^\circ$, while keeping elevation separation at 15$^\circ$. As UEs become more closely spaced, the expected ZF SNR suffers from noise enhancements due to similar channels and maximization of the trace terms referred to in Remark 4. Even separations of 10$^\circ$ are insufficient to approach the performance of uncorrelated channels. The separations needed for achieving uncorrelated channels are unlikely to be seen in reality, due to the width of the required angular separations. Naturally, this is sensitive to the type of multipath environment. Secondly, we can observe that our derived expression of \eqref{zfanalysis14} approximates the simulated expected ZF SNR extremely well, as it remains tight for both correlated and uncorrelated scenarios and across all operating SNRs. The bottom sub-figure depicts the impact of reduced elevation separations, where one can notice a smaller difference in the expected ZF SNR, as the elevation UE separation is reduced from 15$^\circ$ to 10$^\circ$. To this end, UE separation reduction in the elevation does not have as prominent impact as in the azimuth. This is a subject of further study and investigations.  
\begin{figure}[!t]
    \centering
    \includegraphics[width=8.2cm]{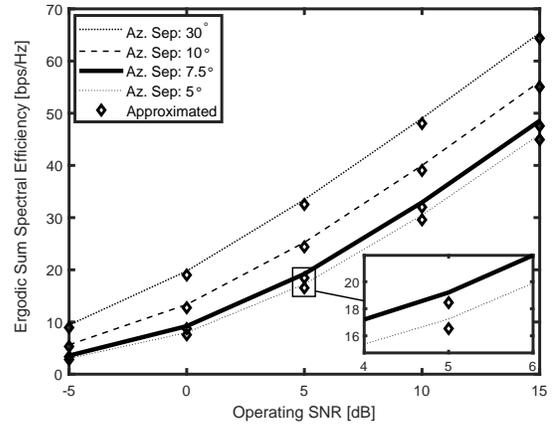}
    \vspace{-9pt}
    \caption{Ergodic sum spectral efficiency vs. operating SNR with $M$=128, $L$=4 and different azimuth UE separations. Elevation separations are 15$^{\circ}$.}
    \label{fig:ErgodicSumSpectralEffvsSNR}
    \vspace{-15pt}
\end{figure}

Figure~\ref{fig:ErgodicSumSpectralEffvsSNR} shows the ergodic sum spectral efficiency performance as a function of the operating SNRs. The exact same simulation scenarios and parameters are employed as for the top sub-figure of  Fig.~\ref{fig:ExpectedZFSNRvsSNR}. We can observe similar trends to those for top sub-figure of Fig.~\ref{fig:ExpectedZFSNRvsSNR}, where the ergodic sum spectral efficiency exhibits an almost linear growth with increasing operating SNR levels. Here also for each scenario, our derived expression in \eqref{zfanalysis15} closely matches the simulated performance.

\vspace{-5pt}
\section{Conclusions}
\label{Conclusions}
\vspace{-2pt}
We articulate a simple, tight and widely applicable analytical framework for understanding the performance of ZF precoding in full-dimensional massive MIMO systems. Via an order two Neumann series, we approximate the expected per-UE ZF SNR and ergodic sum spectral efficiency. Our framework caters for finite multipath propagation channels, correlated and uncorrelated UE positions, and multiple BS array configurations. Our analysis reveals several important insights and describes the interaction of propagation and system parameters on ZF performance. To achieve optimal performance, we show that UEs could be separated in the azimuth by 30$^\circ$ in azimuth and $15^\circ$ in elevation, yet this depends on the underlying multipath conditions. Nevertheless, this will be difficult to achieve in practice.


\end{document}